# Diagnostics of the Synchronization of Self-Oscillatory Systems by an External Force with Varying Frequency with the Use of Wavelet Analysis

A. A. Koronovskii, V. I. Ponomarenko, M. D. Prokhorov, and A. E. Hramov

**Abstract**—A diagnostics method based on a continuous wavelet transform is proposed. This method makes it possible to diagnose the presence of synchronization of the oscillations of a self-excited oscillator locked by an external force with a linearly modulated frequency and to distinguish such a situation from the case when an external signal leaks into self-oscillations; i.e., the signals are summed without a change in the self-oscillation frequency. The method's efficiency is shown with the use of a Van der Pol generator and experimental physiological data as examples.

## INTRODUCTION

It is well known that interaction between nonlinear oscillatory systems of a different nature, including those exhibiting chaotic behavior, may result in the synchronization of these systems [1]. Studies of synchronization in live organisms whose vital functions are governed by the interaction between a large number of complex rhythmic processes have been the focus of attention in recent years [2–4]. Functioning of the human cardiovascular system is a good example of such interaction between different physiological rhythms. The list of the most important oscillatory processes controlling the dynamics of this system includes the basic cardiac rhythm, breathing, and the process of slow regulation of blood pressure and the cardiac rhythm with a frequency of approximately 0.1 Hz [5]. As a result of this interaction, such rhythms manifest themselves in different signals, including electrocardiograms (ECGs), blood pressure, blood flow rate, and cardiac-rhythm variability (CRV) [6].

Quite recently, it has been found that the base rhythms of the cardiovascular system may be mutually synchronized [7–10]. In addition, studies have shown that the systems that set the basic cardiac rhythm and the rhythm of slow regulation of blood pressure may be considered as self-oscillators driven by an external force whose role is played by breathing [10, 11]. This conclusion regarding the character of interaction between the cardiovascular system's rhythms becomes even more convincing if experimental data on breathing of a human, which are characterized by a certain rhythm, are compared with model data obtained in studies of coupled self-oscillators. However, if the breathing rhythm is close to 0.1 Hz, i.e., to the natural frequency of the blood-pressure regulation rhythm, it becomes difficult to distinguish synchronization of these two processes from the respiratory component that is always present in the CRV signal used to analyze slow variations of the blood pressure. Indeed, variability of a human's cardiac rhythm is a result of complex interaction between many physiological processes [12], whose contributions to the CRV are difficult to discern in the case of close frequencies.

In this study, we propose a method based on a wavelet transform that makes it possible to diagnose the presence of synchronization of a self-oscillator's oscillations by an external signal with a linear frequency modulation and to distinguish this situation from the case when an external signal leaks into self-oscillations or, in other words, when the signals are summed without altering the self-oscillation frequency. The method's efficiency is demonstrated on numerical examples and experimental physiological signals.

## 1. STUDIED MODEL

The common character of the phenomena observed in periodically excited self-oscillators of a physical or physiological nature was demonstrated in [7, 9]. The authors of these studies have shown that there is a qualitative similarity between the specific features of synchronization observed in the effect of breathing on heartbeat and the blood-pressure regulation process and in the effect of a periodic force acting on a Van der Pol generator. Let us consider an asymmetrical Van der Pol generator under the action of an external force with linear frequency modulation as an example of interaction

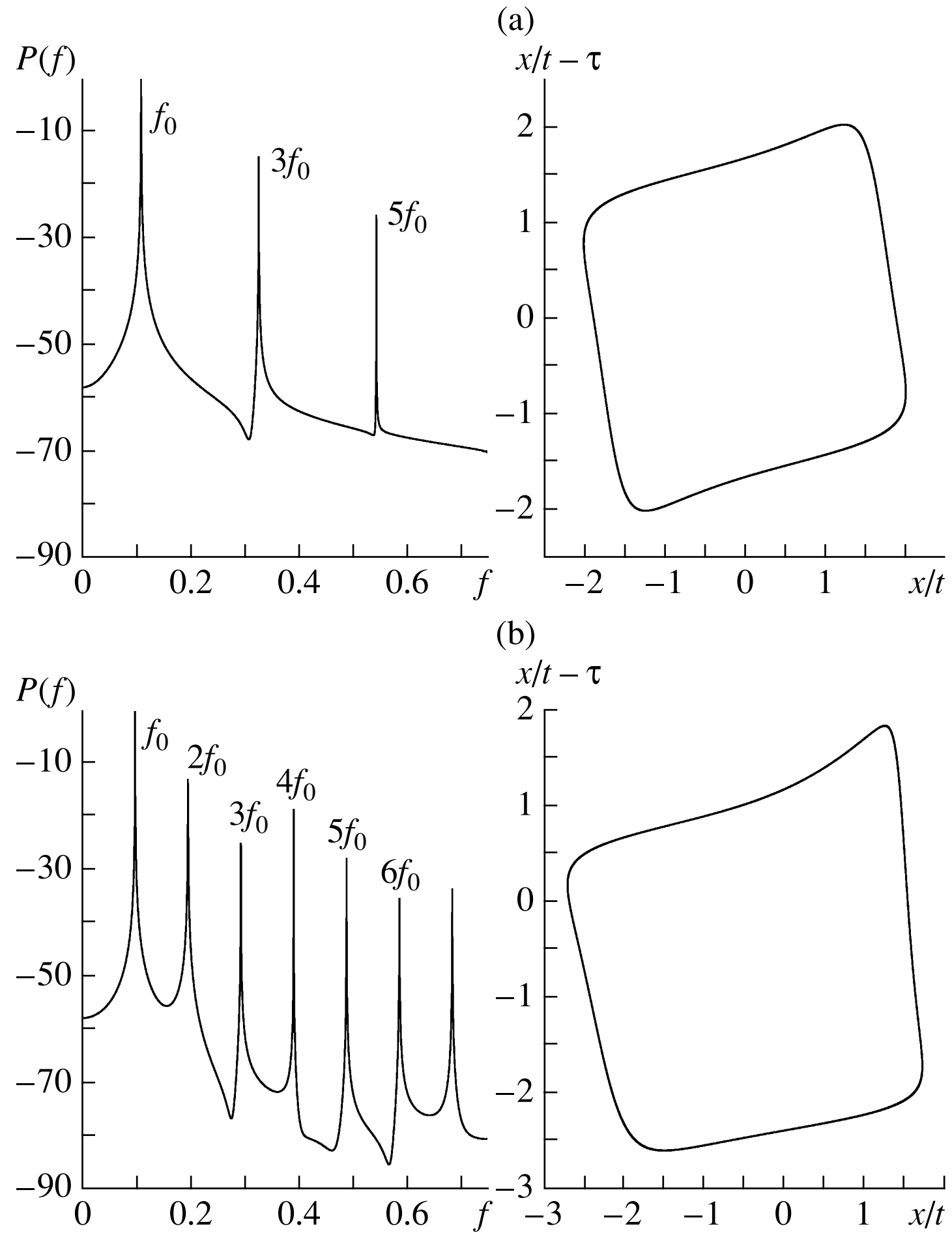

**Fig. 1.** Fourier power spectra (left column) and phase portraits (right column) of oscillations in (a) symmetrical ($\alpha = 0$) and (b) asymmetrical ($\alpha = 1$) autonomous Van der Pol generators.

between breathing and the process of slow regulation of blood pressure.

The studied model of an asymmetrical Van der Pol generator under the influence of a periodic external force is described by the equation

$$\ddot{x} - \mu(1 - \alpha x - x^2)\dot{x} + \omega_0^2 x = K\sin(\omega_L t). \qquad (1)$$

Here, $\mu = 1.0$; $\omega_0 = 0.24\pi$; and frequency $\omega_L$ of the external force depends linearly on time:

$$\omega_L = 2\pi(0.03 + (0.2 - 0.03)t/T), \qquad (2)$$

where $t$ is the current time and $T = 1800$ is the maximum computation time. These values were selected so as to ensure comparison of the results of numerical simulation with the data obtained from the analysis of breathing signals and variability of a human's cardiac rhythm (see Section 4).

Parameter $\alpha = 0$ corresponds to the classical Van der Pol generator characterized by a symmetrical limit cycle. Owing to the symmetry of the phase portrait of oscillations, only odd harmonics of fundamental frequency $f_0$ with frequencies of $(2n + 1)f_0$ ($n = 1, 2, \ldots$) are observed in the power spectrum of oscillations. This situation is shown in Fig. 1a, where the spectrum and the phase portrait of oscillations in a symmetrical Van der Pol generator operating in the autonomous mode are displayed (for the external-signal amplitude $K = 0$). One can see that the power spectrum contains only odd harmonics of the fundamental frequency ($3f_0$, $5f_0$, $7f_0$, etc.). The presence of such harmonics is very inconvenient for simulation of interaction between the basic rhythms of the cardiovascular system since the dynamics of the self-oscillation's second harmonic is clearly exhibited in those rhythms.

Therefore, we propose to use the model of a modified Van der Pol generator with a quadratic nonlinearity as a model for the analysis of synchronization between physiological processes in the cardiovascular system. Here, we consider model (1) with the parameter of quadratic nonlinearity $\alpha = 1$. Corresponding characteristics of autonomous dynamics are shown for this case in Fig. 1b. One can see that the phase portrait is strongly asymmetrical and that the power spectrum contains both even and odd harmonics of fundamental frequency $f_0$. Thus, introduction of asymmetry into the model allows more accurate description of actual systems.

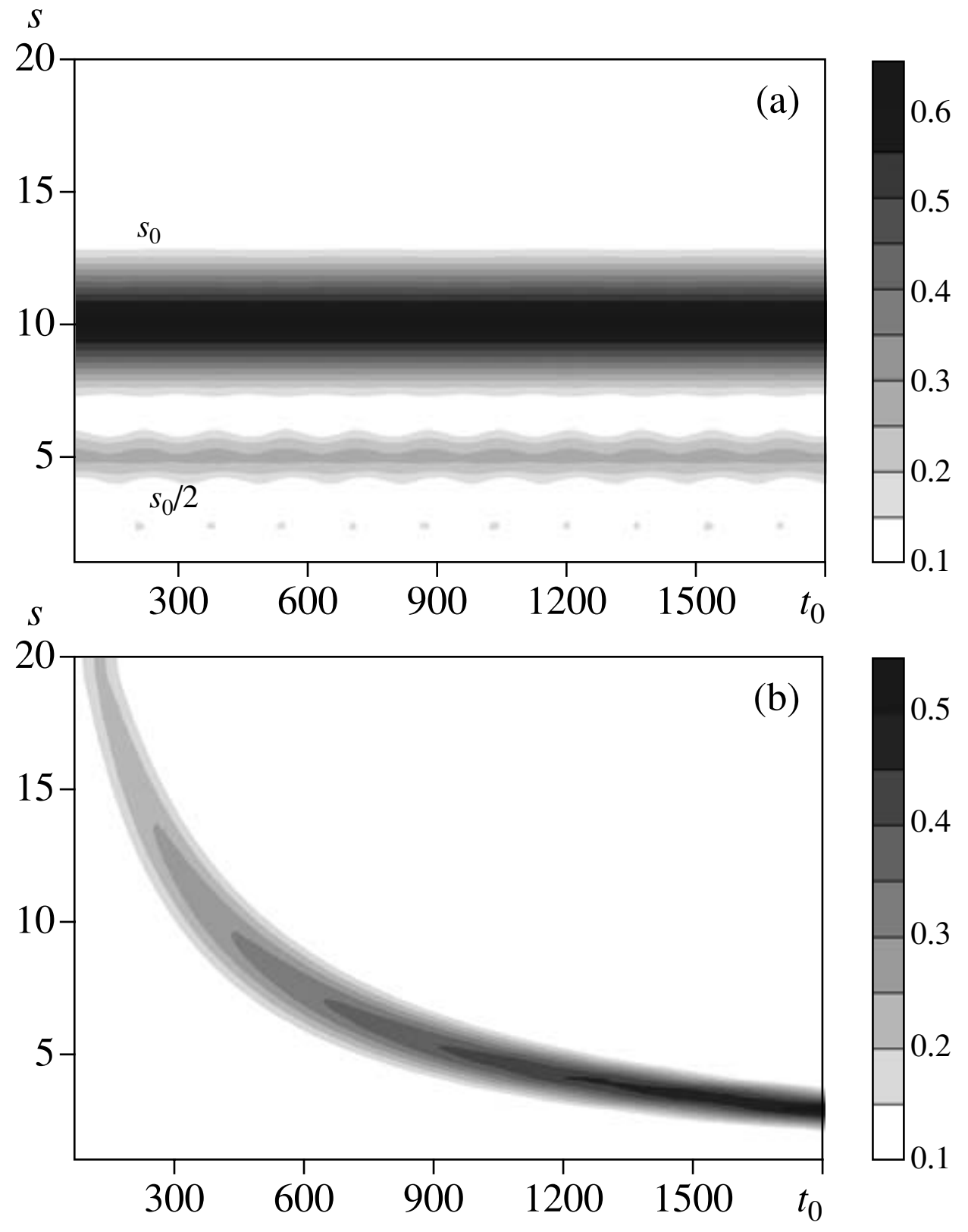

**Fig. 2.** Wavelet spectra $|W(s, t_0)|$ of (a) an autonomous asymmetrical Van der Pol generator for $\alpha = 1$ and $K = 0$ and (b) an external harmonic signal with a variable frequency.

In order to consider and compare situations when oscillations are synchronized by an external signal and an external signal leaks into the experimentally measured signal, we consider not only model (1) but also the sum signal that has the form

$$x_\Sigma(t) = x(t) + R\sin(\omega_L t), \quad (3)$$

where $x(t)$ is a solution of the equation for an autonomous asymmetrical Van der Pol generator, $R$ is the amplitude of the leaking signal, and $\omega_L$ is the frequency of the additive sinusoidal signal described by formula (2).

## 2. APPLYING A WAVELET TRANSFORM TO THE ANALYSIS OF SYNCHRONIZATION OF OSCILLATIONS AND MEASURING SYNCHRONIZATION BETWEEN OSCILLATIONS

During the study of synchronization of chaotic oscillators, different types of synchronous behavior can be distinguished: complete and generalized synchronization, synchronization with a delay, and phase synchronization [1]. In most cases, phase synchronization is considered, for which the phase of the chaotic signal is important (see [1, 13–15]). The onset of phase synchronization means that the phases of chaotic signals are entrained, while the amplitudes of those signals remain uncoupled and look chaotic. The phase entrainment results in the coincidence of the signals' frequencies. The frequency of a chaotic signal is defined as the average rate of change of the signal phase, $\langle \dot{\phi}(t) \rangle$. However, if the signal has a complex spectral composition, it is not always possible to characterize the system by the phase alone [16].

In [17, 18], a new approach to the analysis of synchronization of oscillations was proposed. This approach involves introduction of a continuous set of phases that is defined through a continuous wavelet transform [19, 20] of time series $x(t)$

$$W(s, t_0) = \int_{-\infty}^{+\infty} x(t)\psi^*_{s, t_0}(t)dt, \quad (4)$$

where $\psi_{s, t_0}(t)$ is the wavelet function obtained from mother wavelet $\psi_0(t)$ as

$$\psi_{s, t_0}(t) = \frac{1}{\sqrt{s}}\psi_0\left(\frac{t - t_0}{s}\right). \quad (5)$$

Time scale $s$ controls the width of wavelet $\psi_{s, t_0}(t)$ and $t_0$ is the time shift of the wavelet function along the time axis. (The asterisk denotes complex conjugation.) Note that, during the wavelet analysis, the concept of a "time scale" is commonly used instead of the "frequency" concept typical of the Fourier transform.

The Morlet wavelet $\psi_0(\eta) = (1/\sqrt[4]{\pi})\exp(j\omega_0\eta) \times \exp(-\eta^2/2)$ [21] is used here as the mother wavelet. If the wavelet parameter is selected so that $\omega_0 = 2\pi$, the relation $s = 1/f$ between time scale $s$ of the wavelet transform and frequency $f$ of the Fourier transform is ensured.

The wavelet spectrum

$$W(s, t_0) = |W(s, t_0)|\exp[j\phi_s(t_0)] \quad (6)$$

describes the system's behavior on each time scale $s$ at any time $t_0$. Magnitude $|W(s, t_0)|$ characterizes the presence and strength of corresponding time scale $s$ at time $t_0$.

To visualize 3D wavelet surfaces $|W(s, t_0)|$, we use projections of these surfaces onto the $(s, t_0)$ plane [20]. The intensity of coloring on the projections of the wavelet transform's surfaces is proportional to magnitudes $|W(s, t_0)|$ of the coefficients. Figure 2 shows wavelet spectra $|W(s, t_0)|$ for an autonomous asymmetrical Van der Pol generator (Fig. 2a) and an external harmonic signal with variable frequency (Fig. 2b). It is seen from the wavelet spectrum of the autonomous generator's oscillations that the wavelet spectrum has maxima on the scale $s_0 = 1/f_0$ corresponding to the fundamental frequency and on scale $s_0/2$ corresponding to the second harmonic ($2f_0$) of the fundamental frequency (which are indicated in Fig. 2a). The wavelet spectrum of the variable-frequency signal has a maximum whose position varies with time; this maximum is observed on the scale $s_L(t) = 2\pi/\omega_L(t)$, where $\omega_L(t)$ is set by linear law (2). The growth of the time scale's intensity with time (and, respectively, with an increase in frequency) that can be seen in Fig. 2b is explained as follows: The spectral components of a time series characterized by the same intensity, but different frequencies, are represented by the wavelet transform in the form of maxima with different heights and different widths $\Delta s$ [20]. Therefore, if the frequency increases, the intensity of the corresponding maximum on the wavelet surface increases as well.

Furthermore, it is useful to define an integral distribution of the wavelet-spectrum energy over time scales

$$\langle E(s)\rangle = \int |W(s, t_0)|^2 dt_0. \qquad (7)$$

The phase $\phi_s(t) = \arg W(s, t)$ proves to be defined in a natural way for each time scale $s$. In other words, it is now possible to characterize the behavior of each time scale $s$ through the use of phase $\phi_s(t)$ associated with this scale.

If there is some range of time scales $[s_1, s_2]$ such that, for any time scale $s \in [s_1; s_2]$, the phase entrainment condition

$$|\Delta\phi_s(t)| = |\phi_{s1}(t) - \phi_{s2}(t)| < \text{const} \qquad (8)$$

is met and the fraction of the wavelet spectrum's energy associated with this range is nonzero,

$$E_{\text{sync}} = \int_{s_1}^{s_2} \langle E(s)\rangle \, ds > 0, \qquad (9)$$

then time scales $s \in [s_1; s_2]$ become synchronized and self-oscillators operate under the conditions of synchronization of time scales [17, 18]. In relation (8), $\phi_{s1,2}(t)$ are the continuous phases of the first and second oscillators corresponding to synchronized time scales $s \in [s_1; s_2]$.

Application of such an approach allows not only an efficient analysis of the behavior of a system with an ill-defined phase, where it is impossible to define the continuous phase of a chaotic signal, but also consistent consideration of the chaotic-synchronization phenomenon in its entirety; namely, all types of synchronous behavior of chaotic systems that were known earlier are partial cases of synchronization of time scales (see [18] for details).

Consideration of a continuous set of time scales $s$ and the phases of a chaotic signal associated with these scales and separation of the range of synchronous scales $\Delta s = s_2 - s_1$ allows introduction of a quantitative characteristic of the measure of chaotic synchronization in a nonautonomous system. This measure is defined as a fraction of the wavelet spectrum's energy corresponding to synchronous time scales [17, 18]:

$$\gamma = \int_{s_1}^{s_2} \langle E(s)\rangle ds \Big/ \int_0^{\infty} \langle E(s)\rangle ds, \qquad (10)$$

where $\langle E(s)\rangle$ is the integral distribution that is found for the wavelet spectrum's energy over scales on the basis of formula (7). If $\gamma = 0$, there is no chaotic synchronization of the coupled subsystems. If $\gamma \neq 0$, synchronization of time scales is implemented in the system, thereby implying the existence of synchronized scales for which conditions (8) and (9) are met. If $\gamma = 1$, oscillations in each subsystem are identical or are shifted with respect to each other by some time interval $\Delta\tau$. In the case of synchronization between chaotic oscillators, such an operating mode is called complete or lag synchronization [1]. The growth of $\gamma$ from 0 to 1 indicates that the fraction of energy associated with synchronous time scales $s$ increases.

## 3. RESULTS OF STUDIES OF AN ASYMMETRICAL VAN DER POL GENERATOR UNDER THE EFFECT OF AN EXTERNAL FORCE

### *A. Amplitude Dynamics of the Wavelet Spectra of a Nonautonomous Generator and the Sum Signal*

Let us consider the results of the analysis of wavelet power spectra $|W(s, t)|$ of signal $x(t)$ generated by oscillator (1) synchronized by an external force with variable frequency and sum signal $x_\Sigma(t)$ formed according to formula (3) and containing a sinusoidal signal whose frequency varies according to the same law, i.e., law (2).

Figure 3 shows the wavelet spectra in time shift $t_0$–wavelet-transformation scale $s$ coordinates for two different types of signals plotted in the time interval $T = 1800$. In the same figure, dynamics of the scale $s_L = 2\pi/\omega_L$ corresponding to linearly varying frequency (2) is shown as a dotted curve.

Analysis of the wavelet spectrum in Fig. 3a that corresponds to a generator synchronized by an external signal with a variable frequency shows that, in this case, the classical picture of entrainment of the generator frequency by an external signal is observed. This phenomenon manifests itself in the appearance of kinks located at times $t_s$ and $t_{2s}$ (indicated with arrows in the figure) when the frequency of the external signal is close to either frequency $\omega_L(t_s) \approx 2\pi f_0$ of the autonomous generator or its second harmonic $\omega_L(t_{2s}) \approx 4\pi f_0$ (indicated in the wavelet spectra). This kink reflects the effect of pulling the generator frequency by the external signal and, then, the return of the generator's oscillation frequency (as well as its harmonics) to the free autonomous frequency when the mismatch $(\omega_L - 2\pi f_0)$ becomes large. In the frequency entrainment region, a growth of the amplitudes of the corresponding coefficients in wavelet spectrum $|W(s, t_0)|$ is observed. This growth is determined by the known effect of the growth of the oscillation amplitude in the synchronization beak.

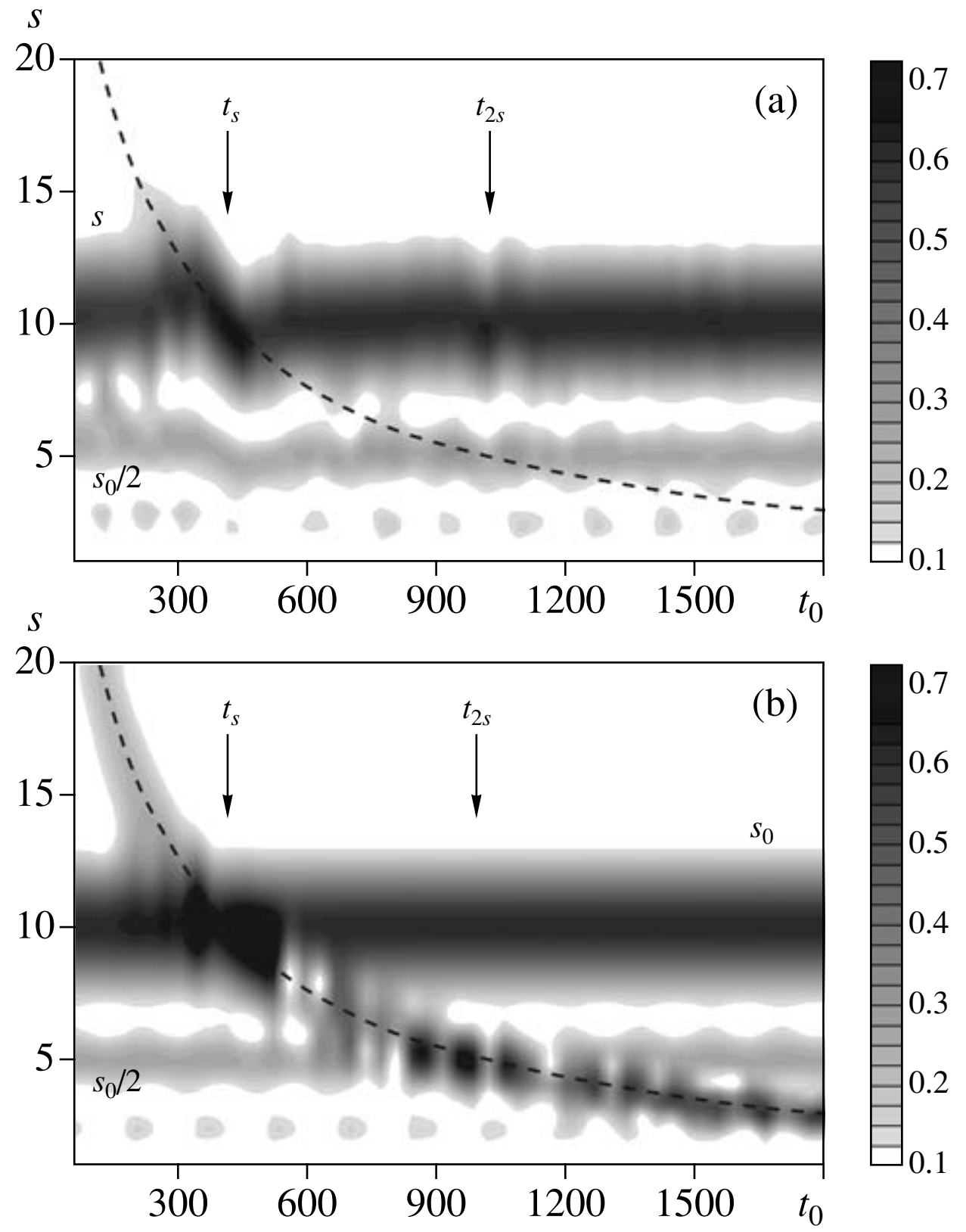

**Fig. 3.** Wavelet power spectra $|W(s, t_0)|$ of (a) a signal generated by oscillator (1) synchronized by an external force with a varying frequency and (b) sum signal $x_\Sigma(t)$ (3) containing a signal with a varying frequency.

It is noteworthy that, in addition to a kink in the spectrum located on main scale $s_0$, a kink is observed on scale $s_0/2$. Note as well that the wavelet surface does not contain maxima corresponding to the external synchronizing signal. This phenomenon is due to the low intensity of the corresponding harmonic.

Quite a different situation is observed during consideration of sum signal $x_\Sigma$ containing a signal with variable frequency. The wavelet power spectrum corresponding to this case is shown in Fig. 3b. Behavior of this wavelet spectrum is substantially different from that of the spectrum in the case of synchronization (Fig. 3a).

First, the spectrum clearly demonstrates both the dynamics of the Van der Pol generator and the dynamics of the leaking signal with variable frequency (compare Fig. 2a), an effect that is not observed in the case of synchronization.

Second, the characteristic pulling of the generation frequency is not observed in this case. In Fig. 3a, this phenomenon manifested itself in the kinks on the wavelet surface located at times $t_s$ and $t_{2s}$. Here, one can observe only that the wavelet surface is slightly deformed and the amplitudes of wavelet coefficients $|W|$ increase at times $t_s$ and $t_{2s}$ as a result of addition of two signals with comparable amplitudes and close frequencies.

Third, distortions of the surface in the region of base scale $s_0$ do not result in any changes in the dynamics on scale $s_0/2$ that corresponds to the second harmonic of signal $x(t)$ of the asymmetrical Van der Pol generator. Similarly, coincidence of frequencies $\omega_L$ and $4\pi f_0$ does not result in the appearance of any changes in the dynamics of the base scale of the wavelet spectrum.

Therefore, an important conclusion can be drawn: An oscillator signal that is rich in the harmonics of the fundamental frequency and synchronized by an external signal can be distinguished from straightforward leakage of a signal. Such a differentiation is made possible through the use of the wavelet power spectrum in which the dynamics of the scales corresponding to the fundamental frequency and its harmonics should be compared. In the case of leakage, no changes whatsoever in the dynamics of the scale whose frequency is close to that of an external signal cause the changes in the dynamics of other characteristic scales (compare the dynamics of scales $s_0$ and $s_0/2$ in Fig. 3b). In the case of synchronization, a characteristic kink in the power spectrum of the wavelet transformation is observed for all characteristic scales (see Fig. 3a).

### *B. Phase Dynamics of the Signal from a Nonautonomous Generator and the Sum Signal*

Let us consider the dynamics of the phases introduced through the continuous wavelet transform of not only signal $x(t)$ (see Section 2) that is generated by oscillator (3) synchronized by an external force with a variable frequency but also the dynamics of sum signal $x_\Sigma(t)$ (3).

We begin with the study of the dynamics of phase difference $\Delta\phi s_L(t)$ between considered signal $x(t)$ (synchronized asymmetrical Van der Pol generator (1)) or sum signal $x_\Sigma(t)$ (determined by formula (3)) and exter-nal signal $R\sin(\omega_L t)$ with a linearly varying frequency. The phase difference will be determined along time scale $s_L(t)$ corresponding to frequency $\omega_L(t)$ (2) of the external signal (i.e., along the dashed lines in Fig. 3). The dependences of phase differences $\Delta\phi_{sL}$ are presented for different cases in Fig. 4.

Let us study the case when an external signal with varying frequency $\omega_L(t)$ leaks into the analyzed signal; i.e., let us consider sum signal $x_\Sigma(t)$ (3). We analyze the phase dynamics in the vicinity of the time $t = t_s$ when the frequency of the external signal is close to the fundamental frequency of the Van der Pol generator: $\omega_L(t_s) \approx 2\pi f_0$. Behavior of the phase corresponding to the base scale $s_0 = 1/f_0$ of oscillations of the Van der Pol generator can be represented as $\phi_{s0}(t) = 2\pi f_0 t + \varphi_1$. The phase of the external signal is $\phi_{sL}(t) = \omega_L(t)t + \varphi_2$, where $\varphi_{1,2}$ are the initial phases. Then, under the assumption that the amplitude of the external leaking signal is significantly smaller than the amplitude of oscillations of the Van der Pol generator and with consideration for the fact that frequency $\omega_L(t) = 2\pi(a + bt)$ linearly increases with time, the phase difference between the two signals forming sum signal $x_\Sigma(t)$ (3) can be presented as

$$\begin{aligned}\Delta\phi_{sL}(t) &= \phi_{s0}(t) - \phi_{sL}(t)\\ &= 2\pi[(f_0 - a)t - bt^2] + \varphi_1 - \varphi_2.\end{aligned} \quad (11)$$

This formula shows that, in the case of a leaking external signal with a linearly varying frequency, the phase difference varies parabolically, the parabola's extremum being located at the time $t = t_s$ when the frequency of the external signal coincides with the autonomous frequency of the generator. The shape of curve $\Delta\phi_{sL}(t)$ proves to be symmetrical with respect to the time $t = t_s$. A similar behavior is observed if the dynamics of the phase difference is considered in the vicinity of scale $s_0 = 1/f_0$ corresponding to the second harmonics of the signal, rather than in the vicinity of base scale $s_0/2 = 1/(2f_0)$.

The described dynamics of phase difference $\Delta\phi_{sL}(t)$ is shown in Fig. 4, where sum signal $x_\Sigma(t)$ (3) is displayed for $R = 0.2$ (curve *1*) and $R = 1.0$ (curve *2*). From the analysis of curves *1* and *2*, we can draw the following conclusions.

First, dependences of the phase difference in the vicinity of times $t_s$ and $t_{2s}$ (for $R = 0.2$) are close to parabolic (see Fig. 4a).

Second, the dependences of this difference are symmetrical with respect to $t = t_s$ in the vicinity of time $t_s$, when frequencies of the autonomous generator and the external signal coincide. This feature is clearly seen in an enlarged image (Fig. 4b). Curve *1* plotted for a small amplitude of the leaking signal $R = 0.2$ (weak leaking of the external signal) has a parabolic shape. Curve *2*, corresponding to the case of strong leakage ($R = 1.0$), likewise has an extremum at the time $t = t_s$; however, this extremum is less pronounced and its shape is distorted as compared to the quadratic curve. This phenomenon is due to the strong effect of the external signal whose amplitude is large and comparable to that of the Van der Pol generator. Third, in the case of strong leakage (curve *2*), the phase difference is constant and close to zero (Fig. 4a) for times that are significantly different from $t = t_s$. This circumstance is due mainly to the fact that the phase difference between the external and sum signals is in fact fixed when the amplitude of the leaking signal is large.

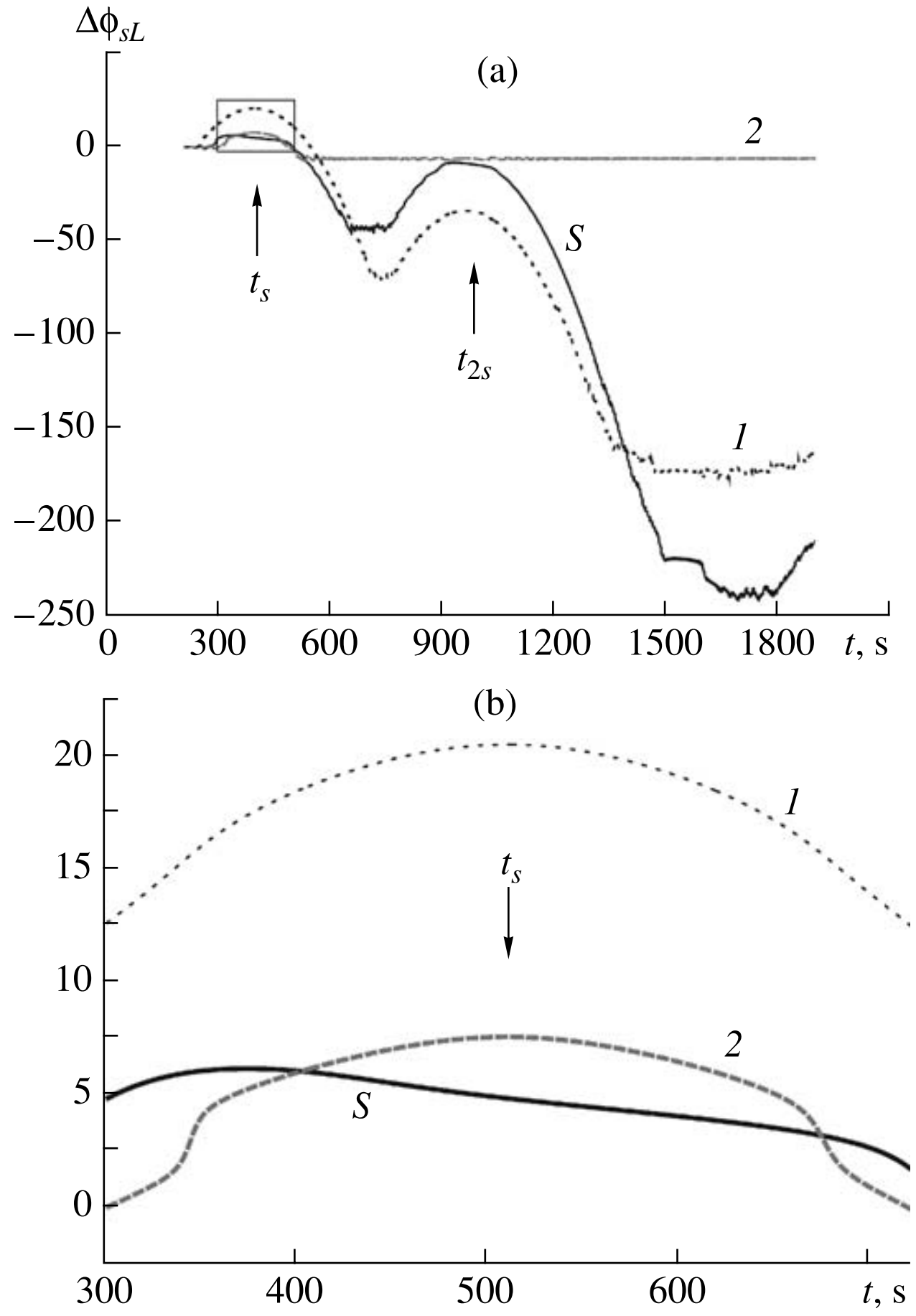

**Fig. 4.** Dynamics of phase difference $\Delta\phi_{sL}(t)$ for the time scales $s_L = 2\pi/\omega_L$ corresponding to linearly growing frequency $\omega_L$ of signal $x(t)$ of an asymmetrical Van der Pol generator and external signal $R\sin(\omega_L t)$. Parameters of sum signal (3) are $R =$ (*1*) 0.2 and (*2*) 1.0. Curve *S* corresponds to the case of synchronization of a Van der Pol generator ($K = 0.2$ in formula (1)). The insert is enlarged in Fig. 4b.

Let us now consider the case of synchronization of asymmetrical Van der Pol generator (1) by an external signal with a linearly increasing frequency. The corresponding dependence of phase difference $\Delta\phi_{sL}(t)$ is shown in Fig. 4 (curve *S*).

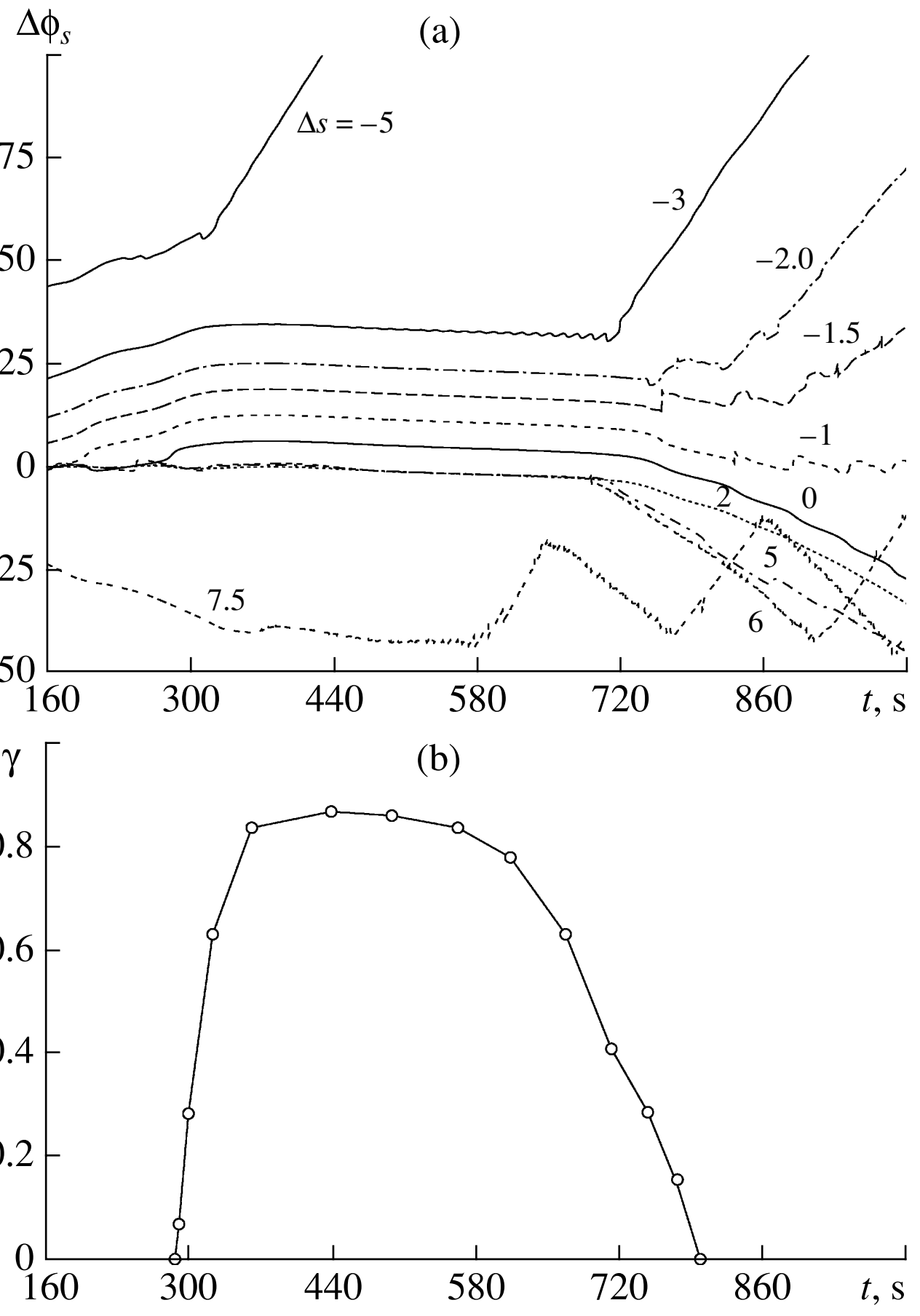

**Fig. 5.** (a) Dynamics of the different phase difference $\Delta\phi_s(t)$ for time scales $s_L(t) + \Delta s$ corresponding to the detuning of linearly growing frequency $\omega_L$ of signals $x(t)$ of an asymmetrical Van der Pol generator and external signal $R\sin(\omega_L t)$. Numbers labeling the curves show the values of $\Delta s$. (b) Synchronization measure $\gamma$ (10) that determines the portion of the wavelet power spectrum's energy associated with the synchronized scales.

Through the analysis of the dynamics of phase difference $\Delta\phi$ between a nonautonomous Van der Pol generator and an external signal by means of the method of slowly varying amplitudes and under the assumption that the external signal does not result in a significant change in the amplitude of nonautonomous oscillations but mainly affects relations between phases, it can be deduced that the value of $\Delta\phi$ satisfies the phase-synchronization equation (the Adler equation [22])

$$d(\Delta\phi)/dt + \kappa\sin\Delta\phi - (2\pi f_0 - \omega_L(t)) = 0, \qquad (12)$$

where $\kappa$ is a factor depending on the generator's parameters. Analysis of the phase-synchronization equation shows that, in the synchronization area set by the conditions $(2\pi f_0 - \omega_L(t)) \le \kappa$, phase difference $\Delta\phi$ becomes $\pm\pi/2$ at the boundary of the synchronization beak and $\pi$ at the center of the synchronization area as the frequency of the external signal changes. This circumstance means that, if a generator is synchronized by an external signal with a linearly varying frequency, it is necessary to fix the time intervals where the phase changes within the $\pi$ range in the vicinity of the times $t = t_s$ and $t = t_{2s}$ (where the frequencies of the first and second harmonics of autonomous oscillations and the external signal coincide).

A similar dynamics of the phase difference is clearly seen in the enlarged fragment (Fig. 4b, curve $S$). Phase difference $\Delta\phi_{sL}$ between the signal of an asymmetrical Van der Pol generator and an external signal with a variable frequency monotonically changes by $\pi$, thereby indicating the presence of synchronization in this case.

Note that a similar phase dynamics is observed at the time $t = t_{2s}$ when the frequency of the external signal is close to that of the second harmonic of the Van der Pol generator: $\omega_L(t_{2s}) \approx 4\pi f_0$.

Therefore, if time dependences $\Delta\phi(t)$ of the phase difference contain segments where $\Delta\phi$ monotonically changes by a value of $\pi$ at times $t_s$ when the frequency of the external signal is close to the frequency of autonomous oscillations (or of its harmonics), the generator's oscillations are synchronized. However, as was noted above, the effect of a leaking signal is characterized by the presence of an extremum of function $\Delta\phi(t)$ at the time $t = t_s$ and the dynamics that is symmetrical with respect to time $t_s$.

Note that application of a signal with a variable frequency is a key element that makes it possible to distinguish between the synchronization of oscillations and the signal leakage. Indeed, if an external signal is characterized by a constant frequency, relation (11) yields that phase difference $\Delta\phi_{sL}(t)$ is time-independent. A similar situation would be observed in the case of "true" synchronization. Therefore, the presence of phase synchronization would be identified in both cases, although, in the first case, phases of the same signal are compared.

Analysis of the dynamics of the phase difference between oscillations corresponding to different time scales $s$ is an important element of the method where synchronization is analyzed by means of a wavelet transform. Let us consider behavior of phase difference $\Delta\phi_s(t)$ determined for the time scales $s(t) = s_L(t) + \Delta s$, where $\Delta s$ is the detuning of time scale $s$ from base scale $s_L(t)$ corresponding to linearly varying frequency $\omega_L(t)$.

The results obtained from the analysis of the phase differences corresponding to different time scales are shown in Fig. 5a in the vicinity of the time interval $t \approx t_s$, where equality $\omega_0 \approx \omega_L$ [roman] holds. Analysis of this figure shows that, if detuning is not large, $\Delta s \in (-1, 2)$, then the dynamics of the phase difference is integrally similar to the case of fine tuning to base time scale $s_L(t)$. If, however, the value of $\Delta s$ is large, the time interval where a synchronous dynamics is observed for the considered scales diminishes and, above some values of detuning $\Delta s$ between the time intervals, a synchronous dynamics is not observed at all.

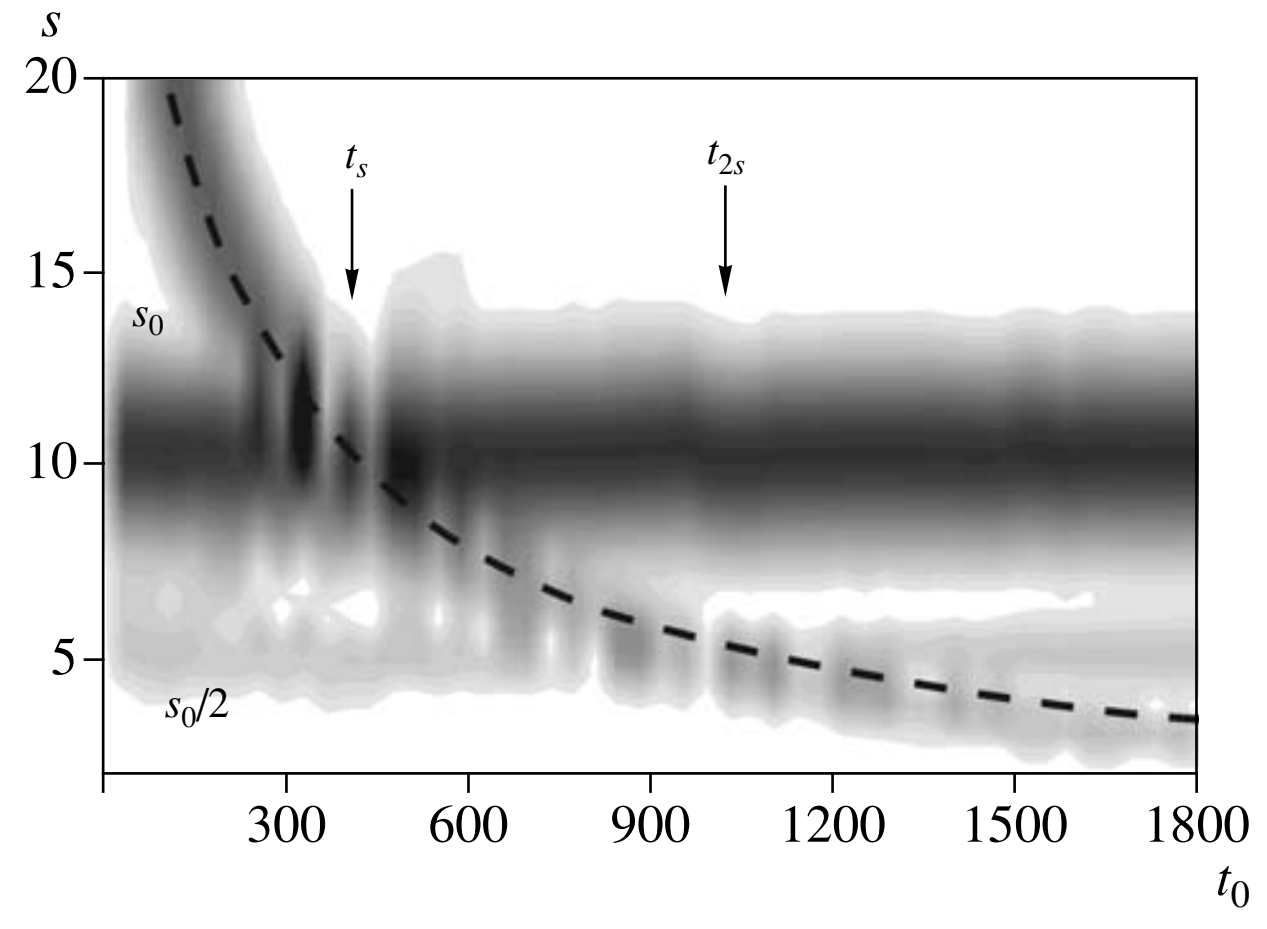

**Fig. 6.** Wavelet power spectrum $|W(s, t_0)|$ of a signal which is a superposition of a signal from generator (1) synchronized by an external force with a varying frequency and amplitude $K = 0.125$ and external signal (3) with the amplitude $R = 1$.

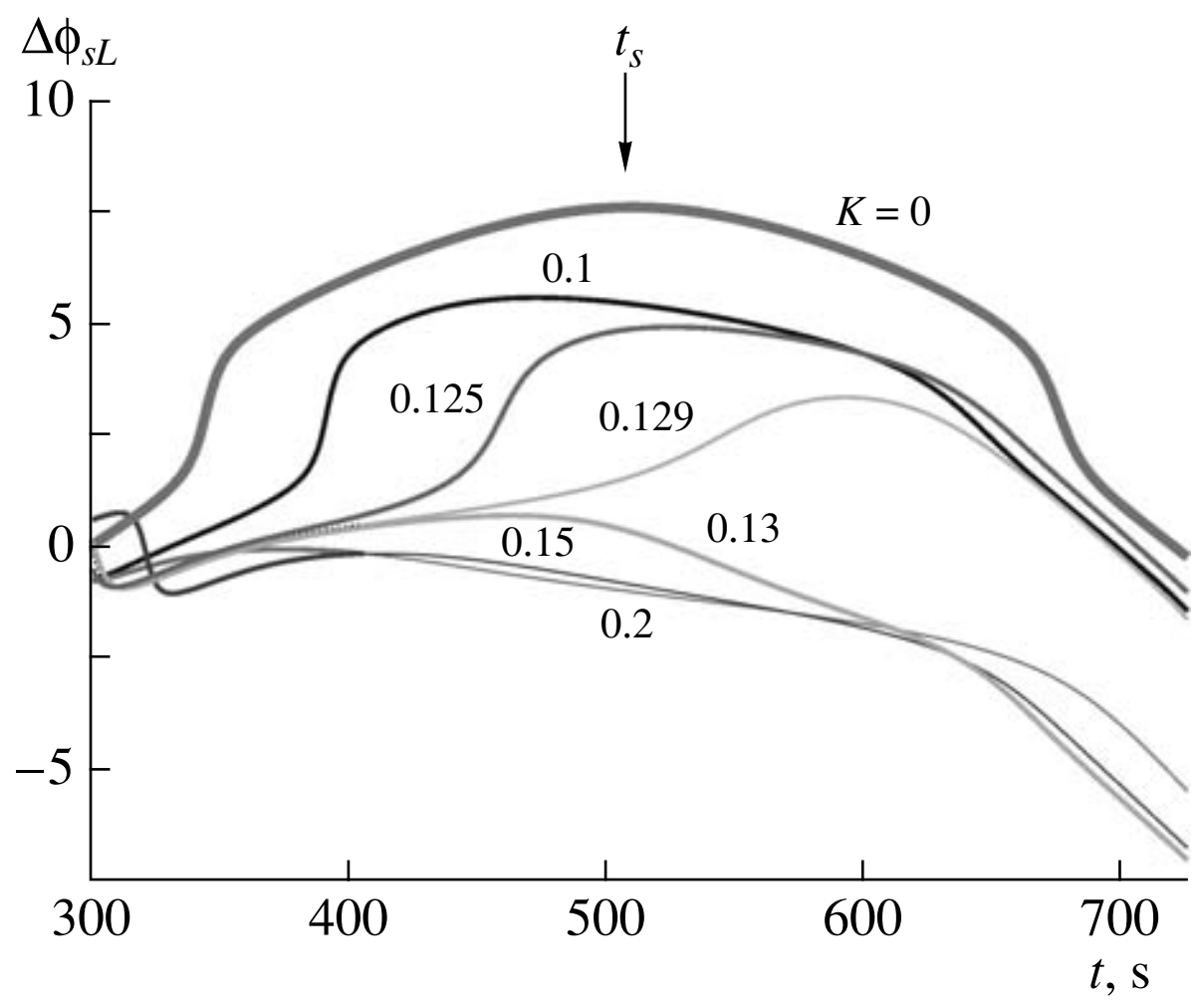

**Fig. 7.** Dynamics of phase difference $\Delta\phi_{sL}(t)$ for time scale $s_L(t)$ (corresponding to linearly growing frequency $\omega_L$) of a signal which is a superposition of a signal from generator (1) synchronized by an external force with a varying frequency and different amplitudes $K$ (indicated by the numbers labeling curves) and external signal (3) with the amplitude $R = 1$.

Synchronization measure $\gamma$ (10) described in Section 2 is an integral characteristic of synchronous behavior of different time scales. Figure 5b shows the time dependence of synchronization measure $\gamma$. One can see that, if time $t \gtrsim 300$, the system rapidly passes to synchronization of the time scales and the portion of the wavelet-spectrum energy associated with the synchronized time scales exceeds 80%. If time increases further, the value of $\gamma(t)$ decreases and vanishes for $t >$ 800 $\gamma$, thus reflecting the absence of synchronization between oscillations.

### *C. Dynamics of a Nonautonomous System in the Case of Concurrent Leakage and Synchronization*

Let us consider the situation when the response of a nonautonomous oscillatory system is characterized by the presence of a leaking signal and synchronization of the self-oscillator. Figure 6 shows amplitude spectrum $|W(s, t_0)|$ of the wavelet transform for a signal representing a superposition of the signal from oscillator (1) synchronized by an external force with a variable frequency and the amplitude $K = 0.125$ and external signal (3) with the amplitude $R = 1$ (compare Figs. 6 and 3). The amplitude spectrum does not allow a definite conclusion to be drawn about whether an external signal leaks into the analyzed series or the self-oscillator's oscillations are synchronized, because the spectrum contains traces of both phenomena. Therefore, to study this issue, it is necessary to consider the phase difference between the external signal with a varying frequency and the signal of the generator's response to an external force.

Figure 7 shows the dynamics of the phase difference for the amplitude of a leaking signal equal to $R = 1$ and different amplitudes $K$ of the synchronizing signal. If $K$ is small, the situation qualitatively corresponds to the case of leaking without synchronization considered in the previous sections. If amplitude $K$ of the synchronizing signal grows, the onset of the synchronization mode is observed in some frequency range. Accordingly, the change in the phase difference within the synchronization beak starts resembling a slanted shelf whose boundaries correspond to loss of synchronization. Outside the synchronization region, an approximately constant phase difference is observed in the case of strong detuning (large times $t$) (not shown in Fig. 7; see Fig. 4a, curve *2*).

Therefore, if leakage is strong and the amplitude of an external synchronizing signal is significant, the change in the phase difference between the external signal and the signal from a Van der Pol generator decreases in the synchronization beak.

## 4. RESULTS OF THE STUDIES OF SYNCHRONIZATION BETWEEN BREATHING AND THE RHYTHM OF SLOW REGULATION OF BLOOD PRESSURE

Let us analyze a physiological time series created by the cardiovascular and respiratory systems of a human. We have studied seven healthy male subjects from 20 to

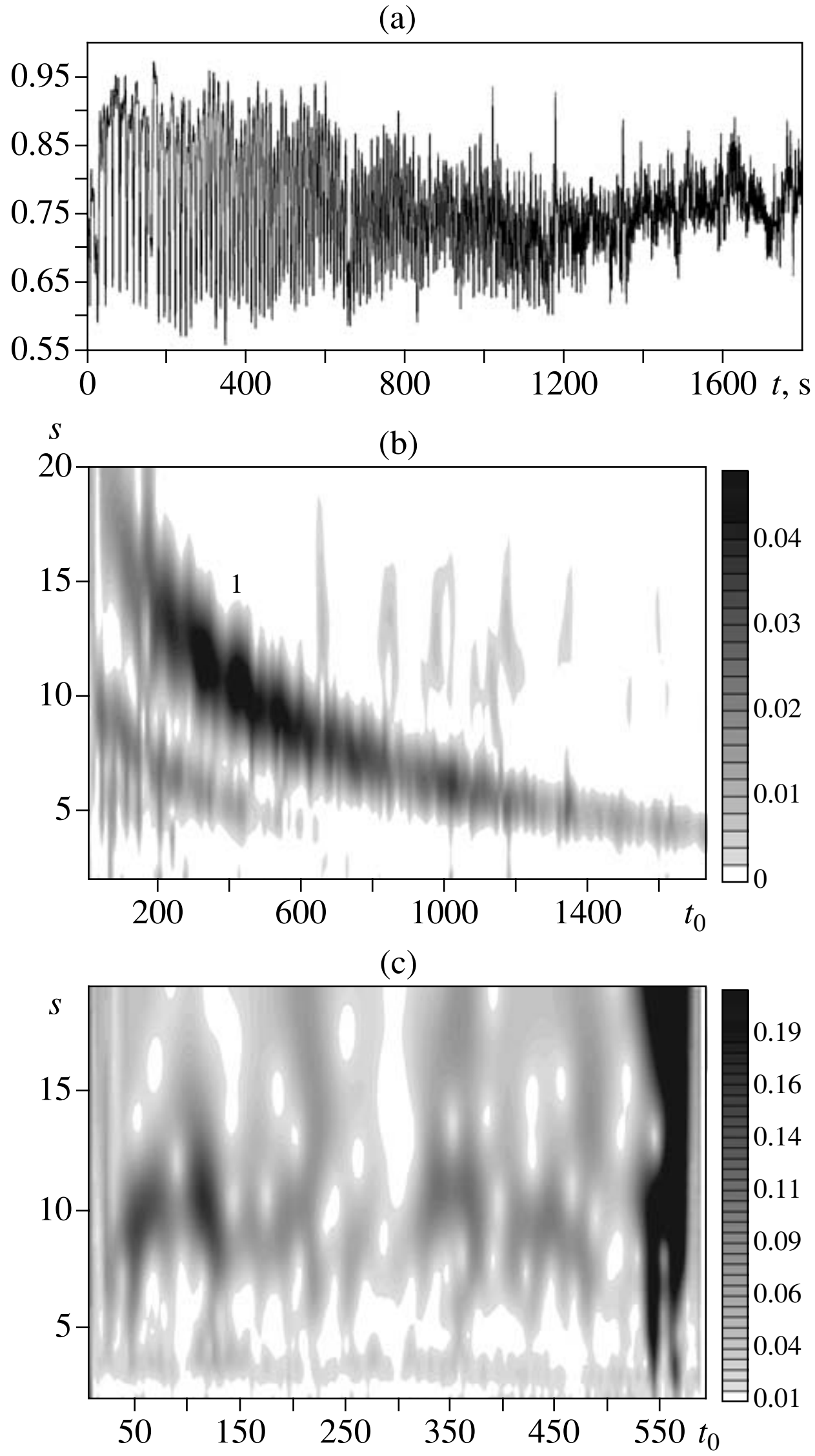

**Fig. 8.** (a) Time series of *R–R* intervals for breathing with a linearly increasing frequency. Wavelet power spectra for (b) a variable breathing frequency of a subject and (c) unrestricted breathing of the same subject.

34 years of age who volunteered to participate in the experiment. All of them exhibited a medium level of physical activity. While a subject was sitting, his ECG and breathing rate were recorded concurrently. All signals were recorded with a frequency of 250 Hz and 16-bit resolution. Afterward, they were processed on a computer.

An experiment has been conducted during which each subject breathed with a set rhythm whose frequency $f_b$ varied from 0.05 to 0.30 Hz. The breathing rate was set by a 0.5-s-long pulsed audio signal. During the signal, a subject inhaled air. No other requirements for breathing were set. Each subject selected the duration of the inhalation and exhalation periods and the respiratory depth that was convenient for him. Breathing with linearly increasing frequency was recorded over 30 min. Before the recording of each specific experiment with a given rhythm started, a subject was given 3 to 5 min to become accustomed to the required breathing rate.

By separating a sequence of *R–R* intervals, i.e., a set of time intervals $T_i$ between two sequential *R*-peaks, in the ECGs, we have obtained information about variability of the cardiac rhythm. Figure 8a shows the analyzed time series of *R–R* intervals obtained for breathing with linearly increasing frequency. Note that time interval $T_i$ between individual counts in the series of *R–R* intervals is not constant; therefore, a technique for performing the wavelet transform of a time series with unequally spaced time counts has been developed. Figure 8b shows wavelet spectra $|W(s, t_0)|$ computed on the basis of *R–R* intervals for the subjects' breathing with linearly increasing frequency and breathing without any restrictions.

Wavelet spectrum $|W(s, t_0)|$ of the time series of *R–R* intervals for the case of breathing with linearly varying frequency (see Fig. 8b) contains a high-amplitude signal corresponding to the variable breathing rate (Fig. 8b, *1*). In addition, the second harmonic of the breathing signal is observed at twice as large a frequency. The power of the rhythm with a frequency of 0.1 Hz (the Mayer wave) is too small to be seen in the figure. Nonetheless, comparison of the wavelet spectrum of *R–R* intervals for the linearly varying breathing rate and the wavelet spectra of a Van der Pol generator shows an analogy between the real and model systems. The shape of the wavelet spectrum of the real system may be interpreted as showing significant leakage of the breathing signal into the response signal without interaction. The breathing-rhythm amplitude is significantly higher than that of the Mayer wave. Other studies (see, for example, [10]) have shown that, if the breathing rate is close to the frequency of the Mayer wave, synchronization between these two rhythms is observed (200–600 s). However, the data on the wavelet spectra do not allow an unambiguous conclusion to be drawn regarding synchronization, although an increase in the variability of the cardiac rhythm near the frequency of the Mayer wave may be considered as an indirect sign of synchronization.

Figure 9a shows the behavior of the phase difference between the breathing signal and the time scale corresponding to this signal in the series of *R–R* intervals. One can see that, for the time scale $s = s_b(t) = 1/f_b(t)$ ($f_b$ is the linearly varying breathing frequency) in the range 200–600 s, the average phase changes can be described by an almost linear relationship (shown with a thick solid line in Fig. 9), thus indicating the presence of the synchronization mode in this frequency range.

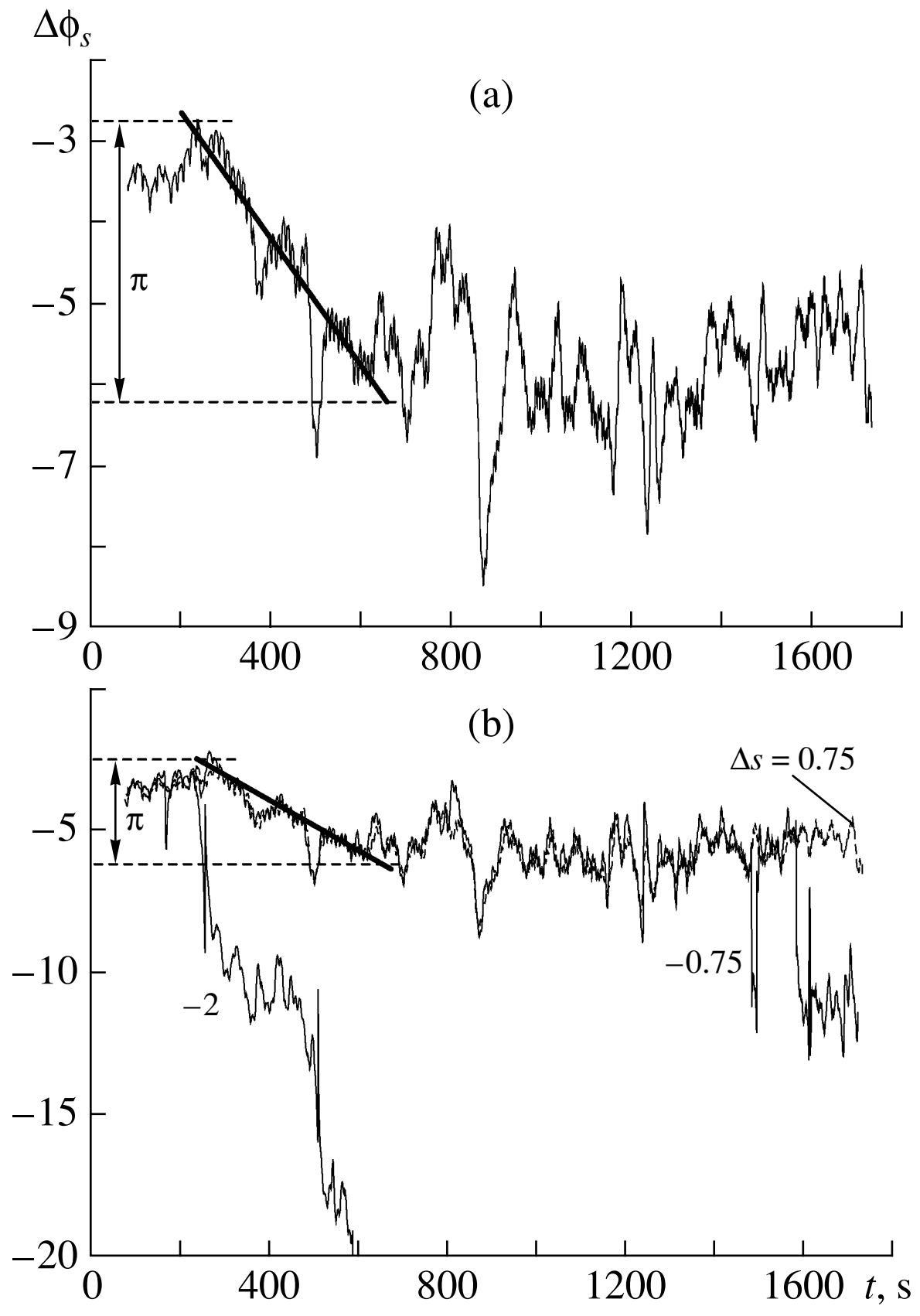

**Fig. 9.** Dynamics of phase difference $\Delta\phi_s(t)$ for (a) linearly changing time scale $s_b(t) = 1/f_b(t)$ and (b) time scales $s_b(t) + \Delta s$ of the series of *R*–*R* intervals and external breathing with a linearly varying frequency. Numbers labeling the curves show the values of $\Delta s$.

Outside the synchronization region, the phase difference oscillates around a constant value. These data allow us to draw a conclusion that, in the range 200–600 s, the breathing dynamics reflected in the signal of the *R*–*R* intervals affects the Mayer wave's internal rhythm, whereas outside this range, the signal of the *R*–*R* intervals shows a regular breathing rhythm. In any case, interaction of this rhythm with other rhythms is not observed.

In addition, the phase difference exhibits similar behavior for adjacent time intervals. Figure 9b shows dynamics of the phase difference for a number of time scales $s = s_b(t) + \Delta s$. One can see that, if the detuning is small ($\Delta s = \pm 0.75$), the dynamics of the phase difference is qualitatively similar to the phase dynamics corresponding to the scale $s = s_b$. Therefore, a small error is permissible during determination of the observation scales and this error will not result in qualitative restructuring of the entire picture. If the difference between the observation time scale and breathing time scale $s_b$ is large, the amplitude of the corresponding rhythm decreases. As a result, it is impossible to reliably determine the phase (see, for example, the dependence corresponding to $\Delta s = -2$ in Fig. 8b), thereby causing the phase difference to cease being constant.

## CONCLUSIONS

In summary, during study of a system where the frequency of an external signal applied to a self-oscillator can be varied, it is possible to distinguish between the situation in which an external force results in synchronization from the situation in which such interaction does not result in synchronization (or, possibly, "circumvents" the system's dynamics and does not produce any effect on it). Moreover, even in the case when leakage occurs, synchronization can be noticed. Unlike the proposed method based on the action on a system with a signal with a varying frequency, the method in which the phase difference is measured at some constant frequency does not provide an answer to the question of whether the system's response results from intense interaction between the rhythms or simply reflects leakage of an external signal into the measured signal without affecting the system's internal dynamics. The use of a wavelet transform for separation of the signal phase allows analysis of the dynamics of the signal phase when the signal frequency is changed. Note that a continuous wavelet transform provides an efficient technique for extraction of the phase from experimental physiological data characterized by high noise levels.

The proposed method can be used in other experiments with physiological systems to diagnose synchronization in the cases when the frequency of the signal applied to the system may be varied.

## ACKNOWLEDGMENTS

This study was supported by the US Civilian Research and Development Foundation (grant REC-006), the Russian Foundation for Basic Research (project nos. 07-02-00747 and 07-02-00044), and the Presidential Council for Grants (the program for the state support of leading scientific schools in the Russian Federation, grant no. NSh-4167.2006.2).

## REFERENCES

1. A. Pikovsky, M. Rosenblum, and J. Kurths, *Synchronization: A Universal Concept in Nonlinear Sciences* (Cambridge Univ. Press, Cambridge, 2001).
2. L. Glass and M. C. Mackey, *From Clocks To Chaos: The Rhythms of Life* (Princeton Univ. Press, Princeton, 1988).
3. L. Glass, Nature **410**, 277 (2001).
4. E. Mosekilde, Yu. Maistrenko, and D. Postnov, in *Chaotic Synchronization, Applications to Living Systems* (World Sci., Singapore, 2002), Ser. A, Vol. 42.

5. S. Malpas, Am. J. Physiol. (Heart Circ. Physiol.) **282**, H6 (2002).
6. A. Stefanovska and M. Hoččič, Prog. Theor. Phys. **139** (Suppl.), 270 (2000).
7. C. Schäfer, M. G. Rosenblum, H.-H. Abel, and J. Kurths, Phys. Rev. E **60**, 857 (1999).
8. M. Bračič-Lotrič and A. Stefanovska, Physica A **283**, 451 (2001).
9. N. B. Janson, A. G. Balanov, V. S. Anishchenko, and P. V. E. McClintock, Phys. Rev. E **65**, 036212 (2002).
10. M. D. Prokhorov, V. I. Ponomarenko, V. I. Gridnev, et al., Phys. Rev. E **68**, 041913 (2003).
11. S. Rzeczinski, N. B. Janson, A. G. Balanov, and P. V. E. McClintock, Phys. Rev. E **66**, 051909 (2002).
12. J. B. Bassingthwaighte, L. S. Liebovitch, and B. J. West, *Fractal Physiology* (Oxford Univ. Press, New York, 1994).
13. M. G. Rosenblum, A. S. Pikovsky, and J. Kurths, Phys. Rev. Lett. **76**, 1804 (1996).
14. V. S. Anishchenko and T. E. Vadivasova, Radiotekh. Elektron. (Moscow) **47**, 133 (2002) [J. Commun. Technol. Electron. **47**, 117 (2002)].
15. A. Pikovsky, M. Rosenblum, and J. Kurths, Int. J. Bifurcation and Chaos **10**, 2291 (2000).
16. V. S. Anishchenko and T. E. Vadivasova, Radiotekh. Elektron. (Moscow) **49**, 123 (2004) [J. Commun. Technol. Electron. **49**, 69 (2004)].
17. A. A. Koronovskii and A. E. Hramov, JETP Letters **79**, 316 (2004)].
18. A. E. Hramov and A. A. Koronovskii, Chaos **14**, 603 (2004).
19. *Wavelets in Physics*, Ed. by J. C. Van den Berg (Cambridge Univ. Press, Cambridge, 1998).
20. A. A. Koronovskii and A. E. Hramov, *Continuous Wavelet Analysis and Its Applications* (Fizmatlit, Moscow, 2003).
21. A. Grossman and J. Morlet, J. Math. Anal. **15**, 273 (1984).
22. R. Adler, Proc. IRE **34**, 351 (1949).